# Competition of superconductivity with the structural transition in $Mo_3Sb_7$

G. Z. Ye[1,5], J.-G. Cheng[1,2*], J.-Q. Yan[3,4], J. P. Sun[1], K. Matsubayashi[2,6], T. Yamauchi[2], T. Okada[2], Q. Zhou[5], D. S. Parker[3], B. C. Sales[3], and Y. Uwatoko[2]

[1]Beijing National Laboratory for Condensed Matter Physics and Institute of Physics, Chinese Academy of Sciences, Beijing 100190, China

[2]Institute for Solid State Physics, University of Tokyo, Kashiwa, Chiba 277-8581, Japan

[3]Materials Science and Technology Division, Oak Ridge National Laboratory, Oak Ridge, TN 37831, USA

[4]Department of Materials Science and Engineering, University of Tennessee, Knoxville, TN 37996, USA

[5]School of Physical Science and Astronomy, Yunnan University, Kunming 650091, China

[6]Department of Engineering Science, University of Electro-Communications, Chofu, Tokyo 182-8585, Japan

*E-mail: jgcheng@iphy.ac.cn

## Abstract

Prior to the superconducting transition at $T_c \approx 2.3$ K, $Mo_3Sb_7$ undergoes a symmetry-lowering, cubic-to-tetragonal structural transition at $T_s = 53$ K. We have monitored the pressure dependence of these two transitions by measuring the resistivity of $Mo_3Sb_7$ single crystals under various hydrostatic pressures up to 15 GPa. The application of external pressure enhances $T_c$ but suppresses $T_s$ until $P_c \approx 10$ GPa, above which a pressure-induced first order structural transition takes place and is manifested by the phase coexistence in the pressure range $8 \leq P \leq 12$ GPa. The cubic phase above 12 GPa is also found to be superconducting with a higher $T_c \approx 6$ K that decreases slightly with further increasing pressure. The variations with pressure of $T_c$ and $T_s$ satisfy the Bilbro-McMillan equation, *i.e.* $T_c^n T_s^{1-n}$ = constant, thus suggesting the competition of superconductivity with the structural transition that has been proposed to be accompanied with a spin-gap formation at $T_s$. This scenario is supported by our first-principles calculations which imply the plausible importance of magnetism that competes with the superconductivity in $Mo_3Sb_7$.

PACS numbers：74.62.-c, 74.62.Fj, 74.70.Ad, 74.40.Kb

# Introduction

In recent years, quantum criticality has been considered as a universal organizing principle for several families of unconventional superconductors,[1] including the heavy-Fermion, cuprate, and iron-based high-$T_c$ superconductors, in which the superconducting transition temperature $T_c$ can usually be enhanced by suppressing a competing electronic order in the normal state via chemical doping or the application of high pressure. To unravel the competitive nature of superconductivity with other electronic orders can thus not only clarify the key factors governing $T_c$, but also deepen our understanding on the pairing mechanism for the observed superconductivity. In this work, we have applied this approach to reveal a competitive coexistence of superconductivity with structural transition in $Mo_3Sb_7$ that has been suggested to be a model system to study the interplay between superconductivity, magnetism, and structural transition.

$Mo_3Sb_7$ has been known since 1960s as the only compound in the Mo-Sb binary system.[2] The revival of interest in this compound arises from the recent discovery of superconductivity below $T_c \approx 2.1$ K as well as the promising thermoelectric properties upon proper doping.[3-5] Although much effort has been devoted to clarify the pairing mechanism of superconductivity, there has been no consensus reached so far. Point-contact Andreev-reflection measurements found a strong anisotropy of the superconducting gap parameter $\Delta$ ($\Delta_{max}/\Delta_{min} > 40$), suggesting an unconventional ($s+g$)-wave pairing symmetry.[6,7] In contrast, specific-heat and muon-spin-ration studies on $Mo_3Sb_7$ support a conventional $s$-wave BCS superconductor; however, it remains under debate whether the superconducting state consists of a single, isotropic gap or two different gaps.[8-12] Moreover, $Mo_3Sb_7$ has been suggested as a coexistent superconductor-spin fluctuation system, in which the observed $T_c$ can be explained only after considering the paramagnon effect in the McMillan equation.[13] Tran *et al*.[14] studied the electrical and magnetic properties of polycrystalline $Mo_3Sb_7$ under pressures up to 2.2 GPa in the temperature range 0.4-80 K. They proposed a pressure-induced spin density wave competes with superconductivity. These arguments point to an unconventional nature of the observed superconductivity and might have a deep root in the peculiar normal state.

$Mo_3Sb_7$ crystallizes in the $Ir_3Ge_7$-type cubic structure with space group $Im$-$3m$ at room temperature.[15] The Mo sublattice is characterized by a three-dimensional network of Mo-Mo dumbbells formed by nearest-neighbor (NN) bond of ~ 3 Å. Alternatively, the Mo sublattice can be regarded as Mo6 octahedral cages at the body-centered positions; these octahedral cages are formed by next-nearest-neighbor (NNN) Mo-Mo bond of ~ 4.6 Å and are connected with each other by the NN bond, as shown in Fig. 1(a). Prior to the superconductivity transition, $Mo_3Sb_7$ undergoes a symmetry-lowering, cubic-to-tetragonal structural transition at $T_s \approx 53$ K. Previous studies have suggested that the structural transition at $T_s$ is accompanied with the opening of a spin pseudogap of $\Delta_s/\kappa_B \approx 120$ K, which has been attributed to the formation of Mo-Mo spin-singlet dimers.[16-19] The importance of magnetism on the structural transition has also been highlighted in a recent study on the lightly doped $Mo_3Sb_7$.[20] The fact that the NN Mo-Mo bond length along the *c* axis is about 0.3% shorter than those within the *ab* plane in the tetragonal phase, Fig. 1(b), indicated that the spin-singlet states could occur only along the *c* axis, leaving the remaining conduction 4d electrons of Mo forming the superconductivity states below $T_c$.[21] Such a scenario could make $Mo_3Sb_7$ a rare example where the localized spin-singlet states coexist with superconducting states below $T_c$.[21] This might be responsible for the above-mentioned contradictions about the mechanism of superconductivity. Although the issue of whether there is any correlation between $T_s$ and $T_c$ has been raised before,[16] direct evidence to confirm the competitive nature of superconductivity with the structural transition or the spin-singlet states remains lacking. In addition, with a structural transition and spin-gap formation above the superconductivity transition, $Mo_3Sb_7$ can be also regarded as an interesting system to explore the quantum criticality with an intimated interplay between lattice instability, magnetism, and superconductivity. By utilizing high pressure as a clean tuning knob, we demonstrate in this work the competitive coexistence of superconductivity with the structural transition, which should serve as a constraint when discussing the mechanism of observed superconductivity in $Mo_3Sb_7$.

## Experimental Details

Single crystals of $Mo_3Sb_7$ used in the present study were grown out of Sb flux. Detailed characterizations on the structural transition, physical properties at the normal and superconducting states have been given elsewhere.[22] All high-pressure resistivity measurements were performed with the standard four-probe method in two different cubic-anvil-type apparatus.[23,24] The first one employs a 250-ton hydraulic press to maintain a constant loading force over massive BeCu guide blocks during cooling down to the lowest temperature ~2 K,[23] while the second one is a miniature, clamp-type "Palm" cubic anvil cell,[24] which enables integration with a $^3$He refrigerator. For both cases, the applied uniaxial loading force is converted by a pair of guide blocks to three-axis compression on a cubic solid gasket made of either pyrophyllite or MgO. The sample was immersed in the liquid pressure-transmitting medium contained in a Teflon capsule that was put in the center of the solid gasket. Four gold wires attached on the sample were introduced out of the Teflon capsule and placed to direct contact with the tungsten-carbide or sintered-diamond anvils. The anvil's top sizes of 4 mm and 2.5 mm have been chosen to generate pressures up to 8 GPa and 15 GPa, respectively. For the constant-force apparatus, the pressure was calibrated at room temperature by monitoring the characteristic resistance change of Bismuth (Bi) at 2.55 and 7.7 GPa;[23] for the "Palm" cubic-anvil cell, the pressure after clamping was calibrated at low temperatures by monitoring the superconducting transition temperature of lead (Pb).[25] Our first-principles calculations were performed using the all-electron planewave density functional theory code WIEN2K,[26] in an attempt to understand the superconducting state and the potential relevance of magnetism.

## Results and Discussions

In this work, we have studied three different pieces of $Mo_3Sb_7$ crystals from the same batch; two pieces were measured with the "constant-force" apparatus, while the third one with the "Palm" cubic anvil cell. The ambient-pressure resistivity $\rho(T)$ shown in Fig. 2(a) is consistent with the previously reported data,[16, 22] featured by a quick decrease at $T_s$ before finally dropping to zero resistivity at $T_c$. To better illustrate these transitions, in this following we define $T_s$ as the maximum of $d\rho/dT$ and $T_c$ as the middle point between 10% and 90% drop of resistivity. As shown in Fig. 2(a) and Table 1, these three samples show similar $T_s \approx 46(1)$ K and nearly identical $T_c \approx 2.4(1)$ K, the latter being among the highest $T_c$ ever reported for $Mo_3Sb_7$. It should be noted that similar procedure has been used to define $T_s$ from resistivity,[22] which is slightly lower than that determined directly from the low-temperature structural study or magnetic susceptibility. Nevertheless, it allows us to track down the systematic variation of $T_s$ with pressure. As noted previously, the normal-state $\rho(T)$ of $Mo_3Sb_7$ is featured by a rather small residual resistivity ratio $RRR \equiv \rho(300\ K)/\rho(5\ K) \leq 1.5$. Single crystals used in the present study show slightly higher $RRR$ values between 1.77 and 2.37. In addition, we have also performed fitting to the $\rho(T)$ data between $T_c$ and $T_s$ with a gap function, viz. $\rho(T) = \rho_0 + AT + Bexp(-\Delta_s/\kappa_B T)$. The obtained spin gaps of $\Delta_s/\kappa_B \sim 100$ K are also close to the reported values.[16, 17, 22] The characteristic temperatures and fitting parameters for these three samples are listed in Table 1. These above characterizations thus ensure the samples' quality and we are in the position to present the $\rho(T)$ data under high pressures.

We first loaded the sample #1 in a "constant-force" apparatus equipped with tungsten-carbide anvils (4 mm-top) and measured its resistivity $\rho(T)$ between 2 and 8 GPa. As shown in Fig. 2(a), the $\rho(T)$ in the normal states decreases steadily upon increasing pressure and the anomaly at $T_s$ shifts down to lower temperatures gradually. In addition, the anomaly at $T_s$ changes from a cusp- to a hump-like feature for $P > 5$ GPa. As shown in the top panel of Fig. 2(a), the variation of $T_s$ with pressure can be seen more clearly from the maximum of $d\rho/dT$, whose magnitude also decreases with pressure. In contrast, the superconductivity transition shown in Fig. 2(b) moves up quickly with pressure and reaches about 3.5 K at 7 GPa, where an obvious broad transition is evidenced. Upon further increasing pressure to 8 GPa, the superconducting transition exhibits a two-step feature with the onset temperature over 5 K. The measurements on sample #1 show that the application of external pressure suppresses $T_s$ but enhances $T_c$. Resistivity measurements under higher pressures are needed to verify: (i) whether the two-step superconducting transition at 8 GPa is caused by an extrinsic pressure inhomogeneity or due to an intrinsic two-phase coexistence, and (ii) how will the $T_s$ evolves with pressure or whether a quantum phase transition can be realized by suppressing completely the cubic-to-tetragonal transition at $T_s$?

These questions are addressed after measurements on the sample #2 up to 12 GPa by changing the tungsten-carbide anvils to the sintered-diamond anvils. As shown in Fig. 2(c, d), the following features are noteworthy: (i) the anomaly at $T_s$ decrease continuously and cannot be discerned any more at $P = 12$ GPa; (ii) the two-step superconducting transition is readily observed at $P = 8.4$ GPa, but at 9.6 GPa and above it changes to a single transition that is coincident with the high-temperature drop of 8.4 GPa data; (iii) $T_c$ decreases slightly with further increasing pressure above 9.6 GPa. These observations unambiguously rule out the extrinsic pressure inhomogeneity as the origin for the two-step superconducting transition, pointing to an intrinsic coexistence of two superconducting phases with different $T_c$s. The high-pressure phase above $P_c \approx 10$ GPa should remain cubic down to the lowest temperature and becomes superconductor with a higher $T_c \approx 6$ K than the low-pressure tetragonal phase. All these above observations based on two different samples were further confirmed on the sample #3 measured with a "Palm" cubic anvil cell apparatus up to 15 GPa, as seen in Fig. 2(e, f). In this case, the two-phase coexistence takes place around 10 GPa, presumably due to the slight pressure variations upon cooling for the *clamp-type* "Palm" cubic anvil cell.

Finally, these above results enable us to construct a temperature-pressure (*T-P*) phase diagram for $Mo_3Sb_7$ shown in Fig. 3, which depicts explicitly the evolution of the cubic-to-tetragonal structural transition at $T_s$ and the superconducting transition $T_c$ as a function of pressure. It becomes clear that the tetragonal phase is destabilized by pressure and the cubic phase remains stable down to the lowest temperature at $P \geq 12$ GPa. There is a two-phase coexistent region around $10 \pm 2$ GPa, signaling a first-order character of this pressured-induced phase transition. Within the tetragonal phase at $P < P_c$, the external pressure enhances $T_c$ with a concomitant suppression of $T_s$, which suggests a competing nature of these two transitions in reminiscent of the quantum criticality observed in several unconventional superconductors.[1] The high-pressure cubic phase is also superconducting with higher $T_c \approx 6$ K, which decreases slightly with further increasing pressure. There is a discontinuous jump of $T_c$ from the tetragonal to cubic phase.

Previous studies at ambient pressure have revealed that the cubic-to-tetragonal structural transition at $T_s$ is accompanied with a spin-gap opening,[16-19, 21] which is most likely associated with the Mo-Mo spin-singlet states formed only along the *c* axis because the NN Mo-Mo bond is about 0.3% shorter than those within the *ab* plane in the tetragonal phase as illustrated in Fig. 1(b). Then, the formation of spin-singlet states will produce a gap over part of the Fermi surface (FS) and leave the remaining Mo 4d electrons within the *ab* plane forming the FS responsible for the observed superconductivity below $T_c$. By suppressing the spin-singlet states, the application of external pressure restores the missing region of FS and thus increases the density of states at Fermi energy available for superconductivity, which could result in the enhancement of $T_c$. Such a scenario of competition for states at the Fermi energy can be further verified by the Bilbro-McMillan equation,[27] *viz.* $T_c^n T_s^{1-n} = T_{c0}$, where $n$ is the portion of electronic density of states at the Fermi energy forming the superconducting gap, and $T_{c0}$ is the superconducting transition temperature without high-temperature transition at $T_s$. This relationship was initially developed to account for the competition of superconductivity with the Peierls-like structural transition in A-15 superconductors like $V_3Si$ and $Nb_3Sn$,[27] and was later found to be also applicable in the

Chevrel-phase superconductor $Eu_xMo_6S_8$ [28] and the heavy-fermion superconductor $CeRhIn_5$ [29], involving the competition of superconductivity with charge-density-wave and antiferromagnetic transitions, respectively.

To estimate the value of $n$ as a function of pressure, we resorted to the upper critical field $\mu_0H_{c2}(T)$, whose initial slope $\eta = -\mu_0 dH_{c2}/dT|_{Tc}$ is proportional to the electronic specific-heat coefficient $\gamma$ and thus $N(E_F)$ via the relationship: $\eta = 4.48 \gamma\rho_0$ (T/K) in the dirty limit of BCS superconductors.[30] Such a BCS-type relationship has been used successfully to reproduce the experimental value of $\eta = 1.25$ T/K for $Mo_3Sb_7$ at ambient pressure,[10] even though it remains controversial regarding the exact pairing symmetry of superconductivity.[6-12] Fig. 4(a-d) show the low-temperature $\rho(T)$ of sample #3 under different magnetic fields and pressures. As can be seen, $T_c$ shifts down to lower temperatures with increasing magnetic fields. Here, we defined $T_c$ as the middle point between 10% and 90% drop of resistivity and plotted the upper critical field $\mu_0H_{c2}$ as a function of $T_c$ in Fig.4(e). The initial slope $\eta$ values are readily obtained from the linear fitting to $\mu_0H_{c2}(T_c)$, while extrapolations to $T = 0$ allow us to estimate the zero-temperature upper critical fields $\mu_0H_{c2}(0)$ as 4.0, 5.1, 6.1, 8.9, 8.7, and 8.5 T for $P = 3.5, 6, 8.5, 12, 14,$ and 15 GPa, respectively. It is also noteworthy that $Mo_3Sb_7$ exhibits a relatively large magnetoresistance $MR(\equiv\rho(H)/\rho(0)-1) > 20$ % in the normal state just above $T_c$ for $P < P_c$, whereas the MR becomes negligible for $P > P_c$.

As shown in Fig. 5, $\eta$ increases quickly from 1.25 T/K at ambient pressure to 1.52 T/K at 8.5 GPa within the tetragonal phase, which indicates that $N(E_f)$ increases with pressure. On the other hand, $\eta$ changes slightly in the cubic phase at $P > 10$ GPa. Finally, we obtained $n \equiv \eta/\eta_{12GPa}$ given that the structural transition disappears at 12 GPa. This leads to an $n = 0.8$ at ambient pressure, which indicates that about 20% of the density of states at the Fermi energy is removed below $T_s$ due to the formation of spin-singlet states. From the pressure dependence of $n$ and $T_s$, we can calculate the pressure dependence of $T_c$ according to the above Bilbro-McMillan equation[27]. As shown by the solid curve in Fig. 3, the experimental $T_c$ can be well reproduced by assuming a $T_{c0} = 4.4$ K, which corresponds to an upper limit of $T_c$ without the high-temperature structural transition. However, this value is lower than the observed $T_c \approx 6$ K for the high-pressure cubic phase. This fact is reflected as a jump rather than a smooth change of $T_c$ near the tetragonal-cubic phase boundary in the $T$-$P$ phase diagram in Fig. 3, presumably due to the reinforced electron-phonon coupling in the higher symmetry cubic phase.

We have performed first-principles calculations to further substantiate our experimental findings in the present study. First, our calculations were performed by using the experimental lattice parameters with internal coordinates relaxed until internal forces were less than 2 mRyd/a.u. In this case, the resultant density of states at Fermi level, $N(E_F)$, was found to decrease with increasing pressure. This is consistent with the general expectation of pressure decreasing interatomic distances, increasing atomic wave function overlap, thereby causing higher band-width and lowering average values of $N(E_F)$. In these calculations, the pressure was introduced by using the experimental lattice parameters.

We found that the $N(E_F)$ values (per Rydberg for both spins) of the tetragonal phase at ambient pressure, 4 GPa and 8 GPa are 281.0, 256.6, 251.9, respectively, and is further reduced to 242.4 at 12 GPa in the cubic phase. This theoretical *reduction* of $N(E_F)$ with pressure therefore cannot provide an immediate explanation for the experimentally observed *enhancement* of $T_c$, which we partially attribute to an *increase* of $N(E_F)$. To resolve this discrepancy we have carefully investigated the possible magnetic behavior, which would likely tend to compete with the superconductivity and could also transfer DOS spectral weight away from the Fermi level, thus affecting the observed $N(E_F)$.

In the relaxed structures we find no evidence of magnetism, with all spin-polarized calculations converging to a result identical to the non-spin-polarized calculations. This is consistent with the lack of magnetism in the tetragonal state, *i.e.* a temperature independent magnetic susceptibility below 20 K [22] and a $1/(T_1 T)$ = constant behavior below ~10 K.[18] However, if the experimental internal coordinates are used in the tetragonal state at ambient pressure, we find a ferromagnetic ground state, albeit with small moments (0.025 $\mu_B$ per Mo for the four-fold site and 0.073 $\mu_B$ per Mo for the two-fold site). A similar situation, i.e. the sensitivity of magnetism to small structural changes, is known in parents of the Fe-based superconductors, as found previously by Subedi[31] for the iron chalcogenide, FeTe, although in that case antiferromagnetism was found for the unrelaxed structure.

Here, we find that the non-magnetic $N(E_F)$ in this unrelaxed tetragonal state, 335.6/Ryd-u.c., is some 20 percent higher than that in the relaxed state, and the Mo-site-projected $N(E_F)$ for the twofold Mo site, 35.64/Mo-Ryd, is much larger than the relaxed value of 21.53. These large $N(E_F)$ values therefore lead, via the Stoner criterion, to a substantially increased tendency towards ferromagnetism. Calculations initialized in an Mo-Mo nearest neighbor antiferromagnetic configuration also converged to this ferromagnetic state.

We now combine this ferromagnetic result with previous theoretical work [32] showing a small magnetic moment configuration to be energetically degenerate with a non-magnetic state. It is certainly possible, though clearly unproven here, that there is a weak ferromagnetic behavior at ambient pressure. This would tend to compete with the superconductivity, since a spin-singlet superconductivity in general does not coexist with ferromagnetism.

In addition to the possible *ferro*magnetism inferred from our calculations, there is also the possibility of *nesting*-based magnetism, which could also compete with the superconductivity. In Fig. 6 we depict the band 291 Fermi surface, which from inspection could support a 'nesting'-induced spin-density wave between the parallel faces at the zone corners. Since this Fermi surface accounts, from our calculations for approximately 15 percent of $N(E_F)$, its 'gapping' (*reducing* $N(E_F)$) at ambient pressure, and then *lack* of gapping (thereby *increasing* $N(E_F)$) as pressure is applied, could yield an increase in the specific coefficient with pressure, as we observe experimentally.

These two types of magnetic behavior present possibilities for explaining the observed enhancement of $T_c$ with pressure. It is also possible that there are significant changes in the phonon spectrum, and ultimately electron-phonon coupling, $\lambda$, with pressure, that affect $T_c$. These would need to be rather significant, given the observed increase in $T_c$ from 2.3 K at ambient pressure to ~ 6 K at 12 GPa. Using the McMillan[33] equation, this $T_c$ increase would require an approximate 35% increase in $\lambda$ (from 0.55 to 0.74), assuming a pressure-independent Debye temperature.

## Conclusion

In summary, we have performed a comprehensive high-pressure study on the resistivity of $Mo_3Sb_7$ and mapped out the pressure dependence of the cubic-to-tetragonal structural transition and the superconductivity transition up to 15 GPa. Below 10 GPa, pressure suppresses $T_s$ but enhances $T_c$. A pressure induced first order transition takes place in the pressure range 8-12 GPa, above which the cubic phase is stable in the whole temperature range. The high-pressure cubic phase is also a superconductor with higher $T_c \approx 6$ K but shows a negative pressure dependence. Our results demonstrated unambiguously a competitive nature between superconductivity and the structural transition within the tetragonal phase; the lower $T_c$ in the tetragonal phase at ambient pressure arises from the competition with the spin-gap formation. The jump of $T_c$ in the cubic phase might be attributed to reinforced electron-phonon coupling in the higher symmetry phase.

## Acknowledgments

This work was supported by the National Basic Research Program of China (Grant No. 2014CB921500), the National Science Foundation of China (Grant No. 11574377), the Strategic Priority Research Program and the Key Research Program of Frontier Sciences of the Chinese Academy of Sciences (Grant Nos. XDB07020100, QYZDB-SSW-SLH013), and the Opening Project of Wuhan National High Magnetic Field Center (Grant No. 2015KF22), Huazhong University of Science and Technology. Work at Oak Ridge National Laboratory was supported by the U.S. Department of Energy, Office of Science, Basic Energy Sciences, Materials Science and Engineering Division.

**Table 1**. Characteristic temperatures and fitting parameters to the resistivity data for three $Mo_3Sb_7$ samples measured in this work.

| Sample No. | 1 | 2 | 3 |
|---|---|---|---|
| $RRR\ [\equiv \rho(300\ K)/\rho(5\ K)]$ | 1.77 | 2.37 | 1.94 |
| $T_c$ (K) | 2.37(2) | 2.36(2) | 2.35(2) |
| $T_s$ (K) | 47(1) | 46(1) | 46(1) |
| $\rho_0$ (μΩ cm) | 99.8(1) | 95.0(3) | 79.6(3) |
| $A$ ($10^{-8}$ Ω cm K$^{-1}$) | 4.2(1.2) | 11.7(2) | 6.9(3) |
| $B$ (μΩ cm) | 89.5(5) | 131.4(9) | 6.7.8(5) |
| $\Delta_s/\kappa_B$ (K) | 99(3) | 101(5) | 83(6) |

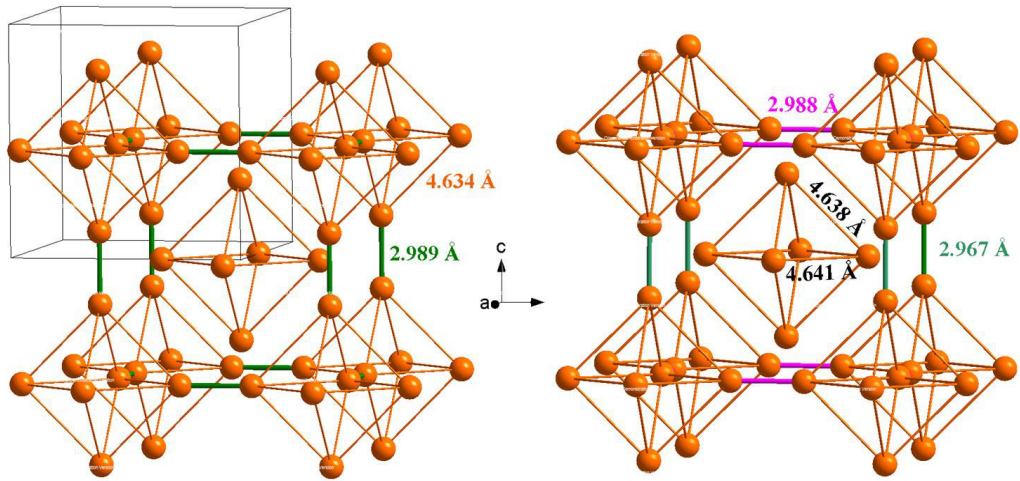

Fig. 1(Color online) The Mo sublattice of $Mo_3Sb_7$ that consists of Mo6 octahedral cages at the body center positions formed by next-nearest-neighbor Mo-Mo bond of ~ 4.6 Å and connected via nearest-neighbor (NN) Mo-Mo bond of ~ 3 Å in the cubic (left) and tetragonal (right) phase. The NN Mo-Mo bond length along the $c$ axis is about 0.3% shorter than those within the $ab$ plane in the tetragonal phase.

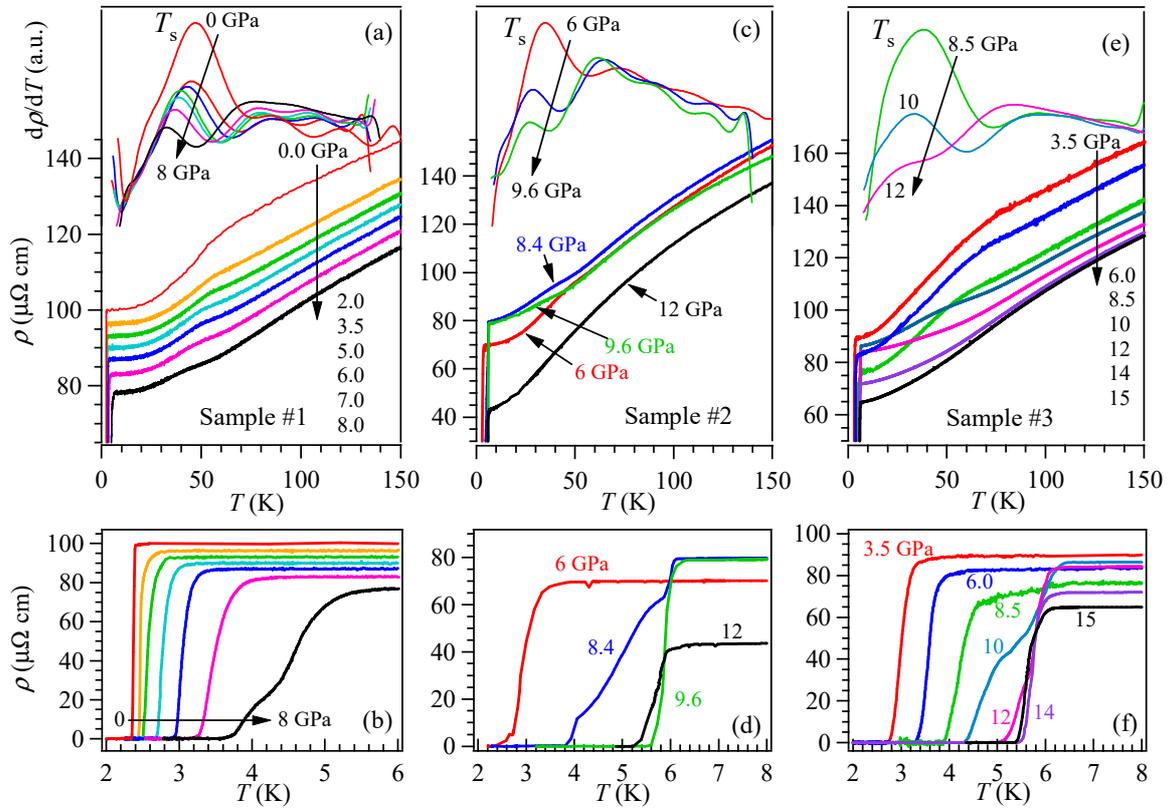

Fig. 2 (Color online) Temperature dependence of resistivity $\rho(T)$ for (a, b) the sample #1 between 2 and 8 GPa, and (c, d) the sample #2 between 6 and 12 GPa, both measured with "constant-force" cubic-anvil-cell apparatus, and (e, f) the sample #3 between 3.5 and 15 GPa measured with "palm" cubic-anvil-cell apparatus. The top panel of (a), (c), and (e) displays the temperature derivative d$\rho$/d$T$ to show the variation of $T_s$ as a function of pressure. Figs. 2(b, d, f) highlight the low-temperature superconducting transition.

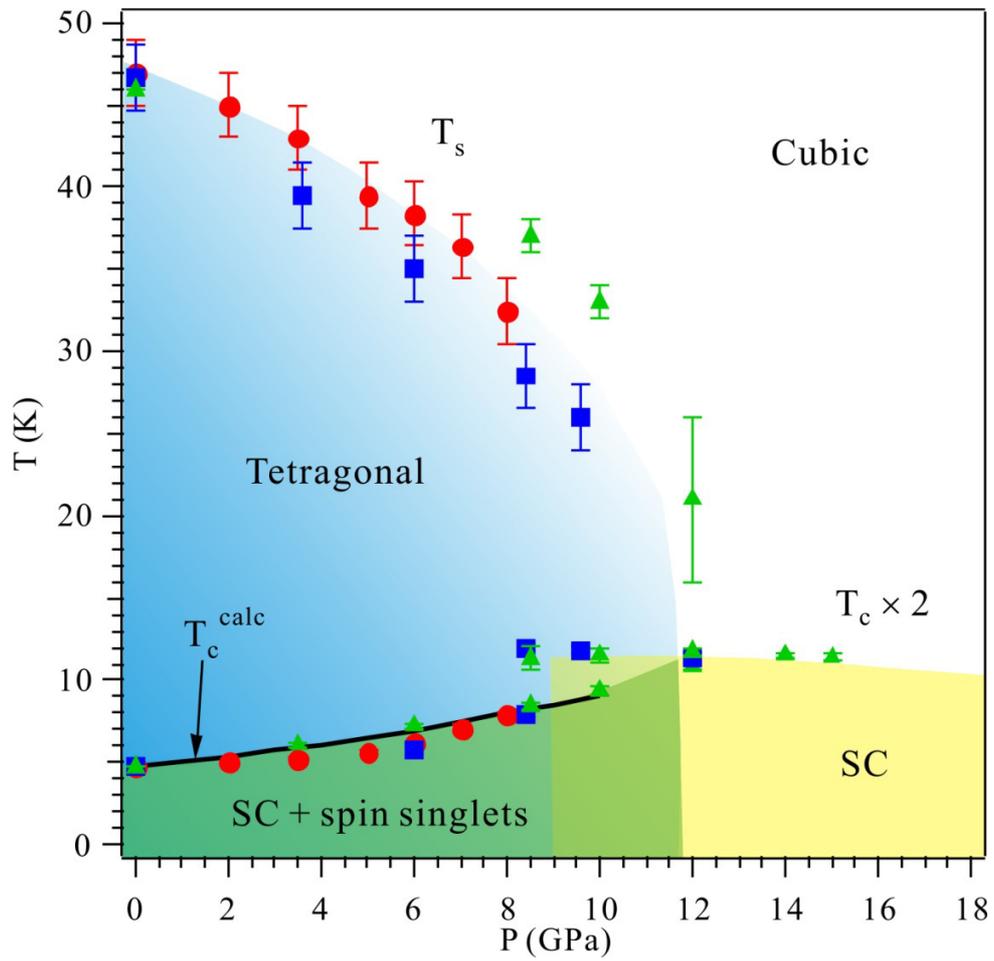

Fig. 3 (Color online) Temperature-pressure phase diagram of Mo$_3$Sb$_7$. The solid black curve denotes the calculated $T_c^{calc}$ according to the Bilbro-McMillan equation as detailed in the main text. The circle, square, and triangle symbols represent the transition temperatures for samples #1, #2, and #3, respectively.

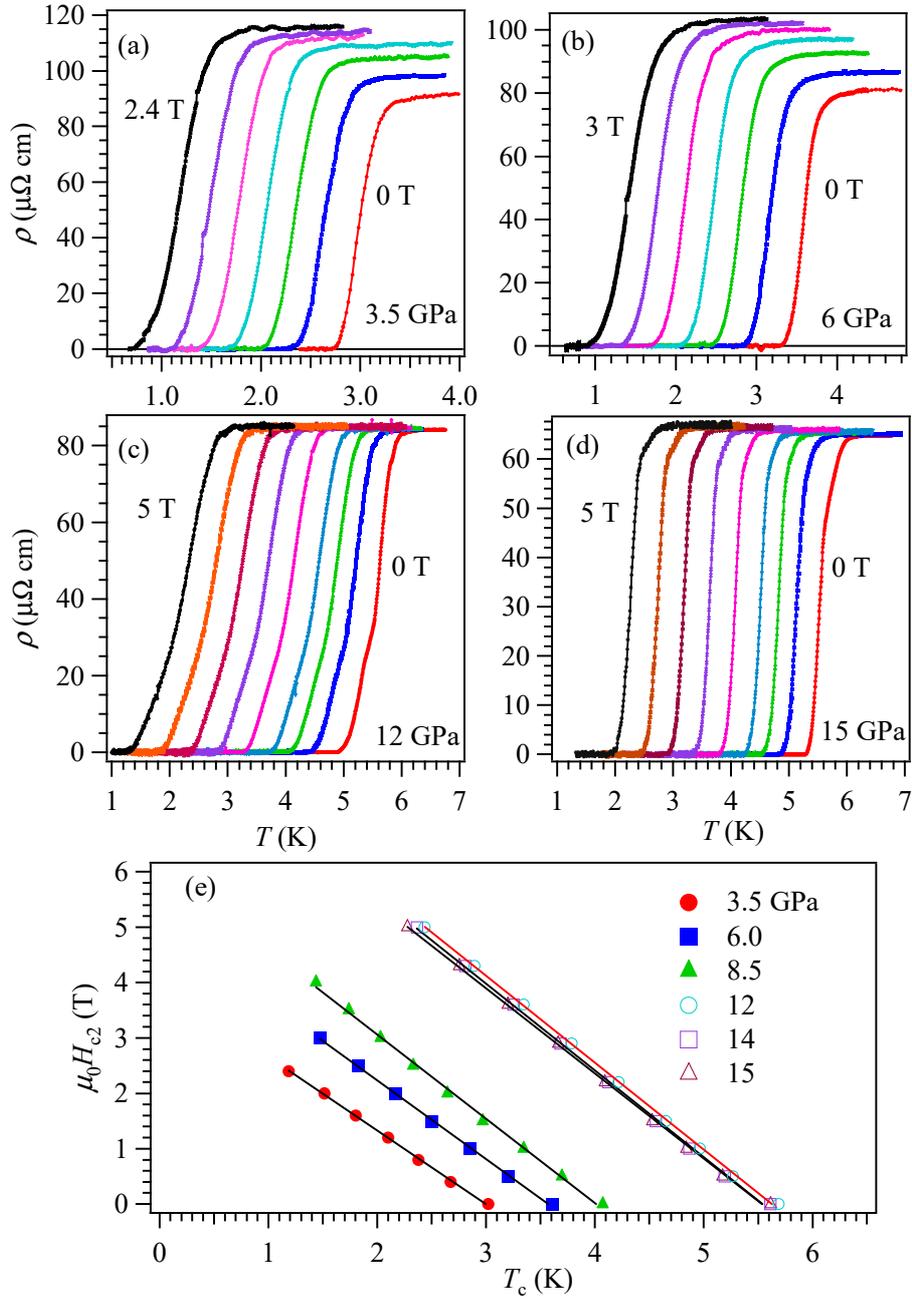

Fig. 4 (Color online) Magnetic-field dependence of the superconducting transition at different pressures: (a) 3.5 GPa, (b) 6.0 GPa, (c) 12 GPa, and (d) 15 GPa. These data were used to obtain the upper critical fields $\mu_0 H_{c2}$ shown in (e), where a linear fitting has been used to extract the values of initial slope $-dH_{c2}/dT|_{T_c}$.

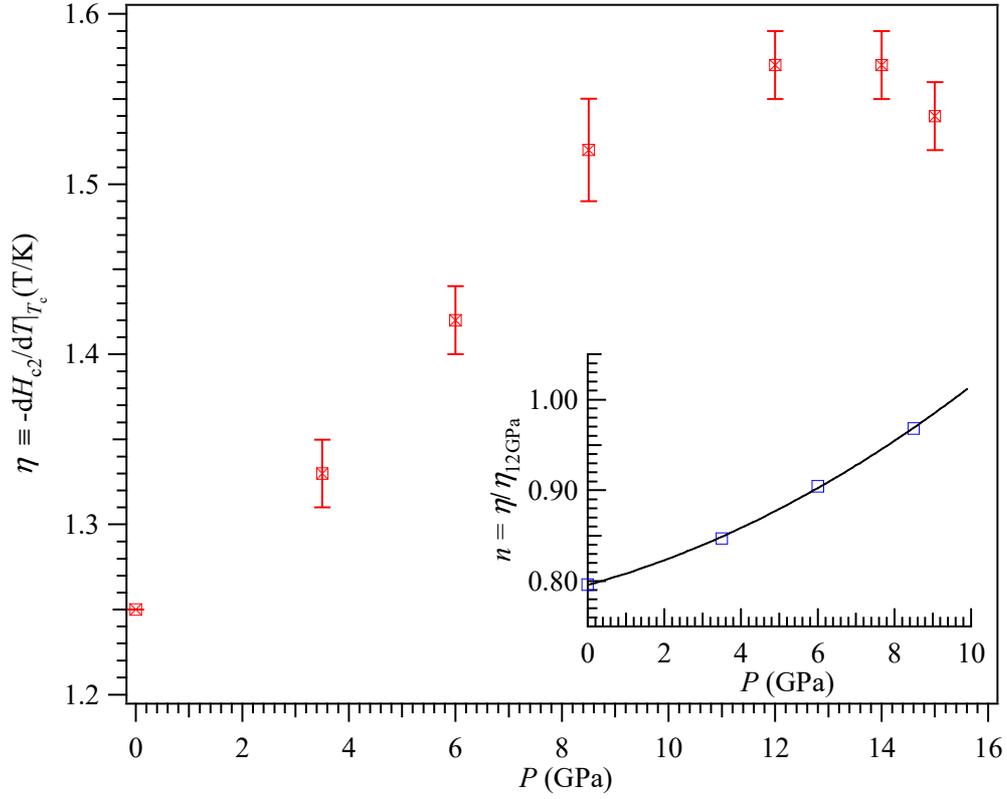

Fig. 5 (color online) Pressure dependence of the initial slope of upper critical field $\eta \equiv -dH_{c2}/dT|_{T_c}$. Since $\eta$ is proportional to electronic specific-heat coefficient $\gamma$ and thus the density of states at Fermi level $N(E_F)$, we employed the obtained $\eta$ values below 10 GPa to estimate the portion of electronic states forming the Fermi surface $n \equiv \eta/\eta_{12GPa}$ as shown in the inset. Here we assume all electronic states at 12 GPa participate the formation of Fermi surface since the spin-singlet states vanish at this pressure.

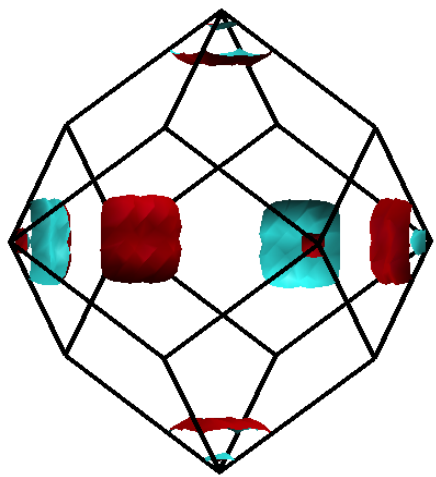

Fig. 6 (Color online) The calculated Fermi surface of band 291 for $Mo_3Sb_7$ in the tetragonal state (the tetragonal distortion is small so that the Brillouin zone depicted is effectively the bcc zone). There is a possibility of nesting between the parallel faces.